\documentclass[a4paper]{article}

\usepackage{INTERSPEECH2021}

\usepackage{amsmath,graphicx}

\usepackage[nolist,nohyperlinks]{acronym}
\usepackage{hhline,subcaption}
\usepackage{gensymb}
\usepackage{xcolor}
\usepackage[hyphens]{url}
\usepackage{amssymb}
\usepackage{comment}
\usepackage{tabularx,colortbl}
\usepackage{multirow}

\usepackage{pifont}% http://ctan.org/pkg/pifont
\newcommand{\cmark}{\ding{51}}%
\newcommand{\xmark}{\ding{55}}%

\newcommand\BlackCell[1]{%
  \multicolumn{1}{c|}{\cellcolor{black}\textcolor{white}{#1}}
}
\newcommand\GrayCell[1]{%
  \multicolumn{1}{c|}{\cellcolor{gray}\textcolor{white}{#1}}
}
\def\arrvline{\hfil\kern\arraycolsep\vline\kern-\arraycolsep\hfilneg}

\acrodef{SNR}{Signal-to-Noise Ratio}
\acrodef{SIR}{Signal-to-Interference Ratio}
\acrodef{ASV}{Automatic Speaker Verification}
\acrodef{TDOAs}{Time Difference Of Arrivals}
\acrodef{VAD}{Voice Activity Detection}
\acrodef{OPR}{Overlapping Ratio}
\acrodef{MCSS}{Multi-Channel Speech Separation}
\acrodef{ICD}{Inter-channel Convolution Difference}
\acrodef{TAP}{Time-Average Pooling}
\acrodef{ASP}{Attentive Statistics Pooling}
\acrodef{MCSV}{Multi-Channel Speaker Verification}
\acrodef{MVDR}{Minimum Variance Distortion-less Response}
\acrodef{WPE}{Weighted Prediction Error}
\acrodef{MFBE}{Mel-Filter Bank Energy}
\acrodef{IRM}{Ideal Ratio Mask}
\acrodef{GEV}{Generalized Eigen-Value}
\acrodef{CNN}{Convolutional Neural Network}
\acrodef{T-F}{Time-Frequency}
\acrodef{GCC}{Generalized Cross-Correlation}
\acrodef{IPD}{Inter-channel Phase Difference}
\acrodef{ILD}{Inter-channel Level Difference}
\acrodef{DFT}{Discrete Fourier Transform}
\acrodef{RIR}{Room Impulse Response}
\acrodef{DS}{Delay-and-Sum}
\acrodef{ASR}{Automatic Speech Recognition}
\acrodef{DOA}{Direction-of-Arrival}
\acrodef{CE}{Cross Entropy}
\acrodef{BCE}{Binary Cross Entropy}
\acrodef{LMFB}{Log-Mel Filter Bank}
\acrodef{STFT}{Short-Time Fourier Transform}
\acrodef{LPS}{Log Power Spectrum}
\acrodef{DNN}{Deep Neural Network}
\acrodef{SNRs}{Signal-to-Noise Ratios}
\acrodef{SIRs}{Signal-to-Interference Ratios}
\acrodef{EER}{Equal Error Rate}
\acrodef{minDCF}{minimum Decision Cost Function}
\acrodef{LR}{Learning Rate}
\acrodef{BLSTM}{Bi-Directional Long Short Term Memory}
\acrodef{PLDA}{Probabilistic Linear Discriminant Analysis}
\acrodef{AF}{Angle Features}
\acrodef{MAE}{Mean Absolute Error}
\acrodef{DPR}{Directional Power Ratio}
\acrodef{TCN}{Temporal Convolutional Network}

\title{Multi-Channel Speaker Verification for Single and Multi-talker Speech}
\name{Saurabh Kataria$^{1\dagger*}$, Shi-Xiong Zhang$^{2\dagger}$, Dong Yu$^2$ \thanks{$^*$This work was done while Saurabh Kataria was a research intern at Tencent AI Lab, USA. $\dagger$These authors contributed equally.}}
\address{
$^1$Center for Language and Speech Processing, Johns Hopkins University, Baltimore, MD, USA\\
$^2$Tencent AI Lab, Bellevue, WA, USA}
\email{skatari1@jh.edu, auszhang@tencent.com, dyu@tencent.com}

\begin{document}

\maketitle
\begin{abstract}
To improve speaker verification in real scenarios with interference speakers, noise, and reverberation, we propose to bring together advancements made in multi-channel speech features.
Specifically, we combine \emph{spectral}, \emph{spatial}, and \emph{directional} features, which includes inter-channel phase difference, multi-channel \emph{sinc} convolutions, directional power ratio features, and angle features. To maximally leverage supervised learning, our framework is also equipped with multi-channel speech enhancement and voice activity detection. On all simulated, replayed, and real recordings, we observe large and consistent improvements at various degradation levels. On real recordings of multi-talker speech, we achieve a 36\% relative reduction in equal error rate w.r.t. single-channel baseline. We find the improvements from speaker-dependent \emph{directional} features more consistent in multi-talker conditions than clean. Lastly, we investigate if the learned multi-channel speaker embedding space can be made more discriminative through a contrastive loss-based fine-tuning. With a simple choice of Triplet loss, we observe a further 8.3\% relative reduction in EER.
\end{abstract}
\noindent\textbf{Index Terms}: multi-channel speaker verification, multi-talker, overlapped speech, speech separation, joint learning

\section{Introduction}
Speech devices are getting equipped with multi-channel and multi-modal information, in turn improving spatial ambiguity and directivity~\cite{chen2018multi,gu2020multi}.
Such information is fused at various levels: feature, embedding, or score~\cite{movsnerfar}.
This is shown beneficial for Automatic Speech Recognition~\cite{sainath2017multichannel}, Speaker Recognition~\cite{cai2019multi}, Speech Enhancement~\cite{gu2020enhancing}, and Source Separation~\cite{wang2018combining}.

Several \emph{spectral}, \emph{spatial}, and \emph{directional} features~\cite{chen2018multi} are proposed for multi-channel version of such problems.
In \cite{liu2020multichannel}, authors use multi-channel \emph{sinc} convolution filters and show the importance of phase for time-domain multi-channel speech enhancement.
Source Separation work of \cite{wang2018multi} proposed multi-channel Deep Clustering which employs \ac{IPD} and asserts that spatial features are helpful even for arbitrary mic configurations.
\cite{gu2020enhancing} improved over \ac{IPD} features through \ac{ICD} features by directly learning on multi-channel temporal data.
In \cite{subramanian2020far}, authors showed the effectiveness of \ac{AF}~\cite{wang2018spatial} for targeted speech extraction.
%For multi-channel dereverberation, \cite{wang2020multi} proposed to use complex STFT features from different mics.
%Most prior multi-channel speech systems pursue pre-processing techniques (like dereverberation), coherence estimation~\cite{yousefian2011dual}, and masking-based beamforming~\cite{luo2020end}.

\ac{MCSV} is an under-explored problem with no common benchmarks except CHiME-5 ~\cite{garcia2019speaker}.
It is pursued mostly using pre-processing techniques like dereverberation and mask-based beamforming.
In \cite{yang2019joint}, authors combined \ac{WPE} dereverberation and \ac{MVDR} beamforming for \ac{MCSV}.
In \cite{taherian2019deep}, authors search for optimal beamformer among variants of \ac{IRM} based \ac{MVDR} and \ac{GEV} beamformers.
\cite{taherian2020robust} showed the utility of multi-channel speech enhancement for \ac{MCSV}.
\cite{cai2019multi} found that multi-channel \ac{T-F} representations are superior to single-channel especially when 3D convolutions are used instead of 2D in a deep multi-channel input based \ac{CNN}.

We believe prior works are not comprehensive in terms of leveraging all available information.
For example, they do not investigate if spatial and location clues of sound sources can help speaker verification.
\cite{wang2018multi} noted that linear beamformer filtering cannot capture multi-channel non-linear information.
Towards building a \emph{complete} multi-channel speaker embedding system, we propose to incorporate various multi-channel features and learn a multi-channel \ac{CNN} embedding network in conjunction with joint (multi-channel) enhancement and \ac{VAD}.
Specifically, we pursue text-independent wide-band \emph{informed}~\cite{delcroix2018single} (opposed to \emph{blind}~\cite{yu2017permutation}) \ac{MCSV}.

Our contributions are:
(1) we provide the first study to compare and combine various multi-channel \emph{spectral}, \emph{spatial}, and \emph{directional} speech features for robust \ac{MCSV}, and report consistent improvements on real far-field data;
 (2) we devise a comprehensive location-aware speaker verification setup with joint enhancement, \ac{VAD}, and realistic single and multi-talker  test scenarios;
 (3) we demonstrate that this multi-channel system performance can be improved via contrastive fine-tuning.
% (3) we show that such improvement can be further extended by fine-tuning the multi-channel system with a simple choice of contrastive loss.

\section{Multi-Channel Speaker Verification}
\subsection{Proposed Multi-Task Supervised Learning Framework}
To maximally exploit supervised learning, we devise a multi-task framework that jointly optimizes three multi-channel tasks: speaker embedding learning, voice activity detection, and speech enhancement.
Thanks to simulation, we utilize ground truth labels for the latter two tasks.
%, as opposed to multi-task formulation of \cite{wang2018combining}.
Using the initial single-channel clean speech, we compute its energy \ac{VAD} and the spectrogram as targets for the two tasks respectively.
Loss for complete framework is $\mathcal{L}_{\text{all}} = \mathcal{L}_{\text{BASE}} + \lambda _{\text{enh}} \mathcal{L}_{\text{enh}}$, where $\mathcal{L}_{\text{BASE}} = \mathcal{L}_{\text{emb}} + \lambda _{\text{VAD}} \mathcal{L}_{\text{VAD}}$, and $\lambda _{\text{VAD}}$, $\lambda _{\text{enh}}$ are regularization weights for \ac{VAD} and enhancement task respectively.
They are set to 0.1 and 0.0005.
$\mathcal{L}_{\text{emb}}$ is \ac{CE} loss,
 $\mathcal{L}_{\text{VAD}}$ is frame-wise \ac{BCE} loss,
 $\mathcal{L}_{\text{enh}}$ is spectrogram-domain \ac{MAE} loss,
and $\mathcal{L}_{\text{BASE}}$ is the loss for the BASE model which is a single-channel joint speaker embedding and \ac{VAD} system without the speech enhancement module.
%$\mathcal{L}_{\text{all}}$ is the loss for the system which learns all tasks.

\begin{figure*}[htbp]
    \centering
    \includegraphics[width=1.0\linewidth]{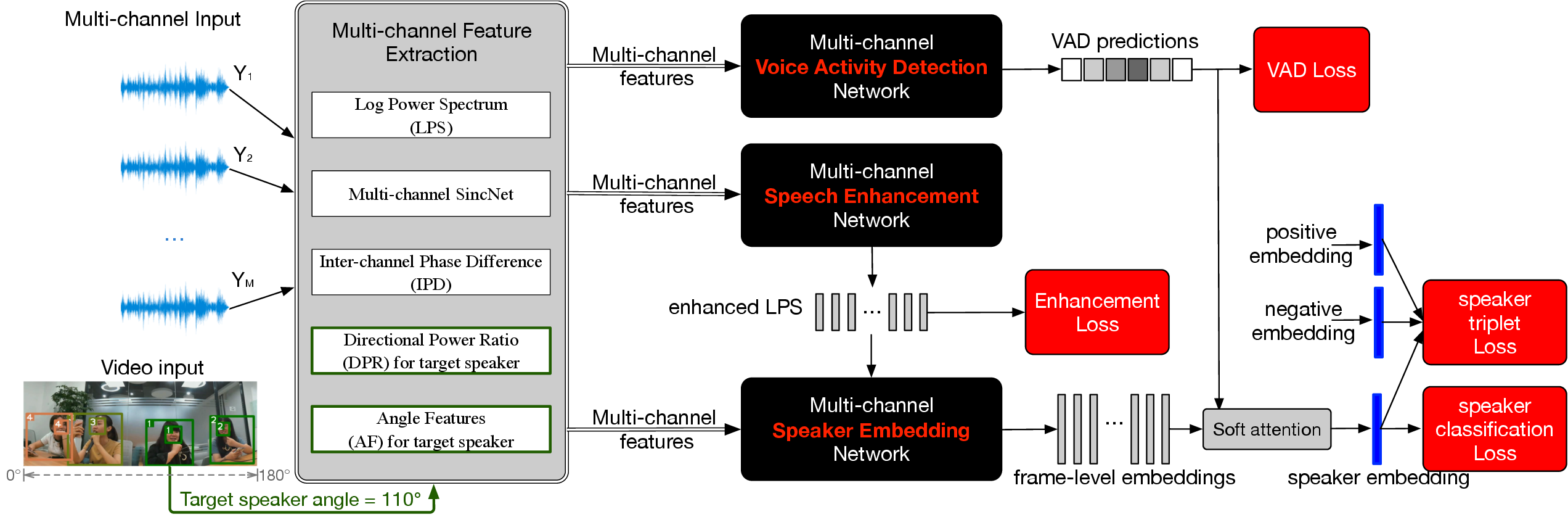}
    \caption{Schematic of the proposed multi-channel speaker verification based on multi-channel features. In practice, face and body detection in video can provide locations of each speaker but not who is speaking or when.}
    %\squeezeup
    %\squeezeup
    %\squeezeup
    %\squeezeup
    \label{fig:multchan} \vspace{-0.5cm}
\end{figure*}

Fig.\ref{fig:multchan} illustrates our proposal.
Note that the input features for the three tasks are identical but their outputs are combined later.
Our methodology is to \emph{incrementally} incorporate various multi-channel features on top of the BASE system in order to maximize speaker recognition performance.
%Such features (explained in Sec. \ref{sec:feats}) fall in the following categories: \emph{spectral} (\ac{LPS}, \ac{LMFB}), \emph{spatial} (\ac{IPD}, sinc convolutions, beamformer, absolute phase), and \emph{directional} (angle features).
We do not investigate feature fusion schemes and simply concatenate all features along the \emph{channel dimension} of the input layer of \ac{CNN}.
We hope to observe the complementary effect of such features~\cite{wang2018multi} since some of them may under-perform in adverse scenarios.
%, especially from \emph{directional} features which improve discriminability by using audio source locations~\cite{gu2020multi}.
For e.g., under heavy reverberations, \ac{IPD}s get degraded.
Hence, we are interested in a \ac{MCSV} system which is robust and generalizes well.
To explore this, we choose a Mandarin test set that is language-mismatched with the train set (English) and is corrupt with reverberations, background noises, and speaker interferences.
Since there is no \ac{MCSV} work directly comparable to our setup, for the baseline, we use the standard single-channel 80-D \ac{LMFB} features (BASE system).

\subsection{Large-Margin Contrastive Fine-Tuning}
We investigate if the learned multi-channel speaker embedding space can be made more discriminative using contrastive loss-based post-processing.
We are inspired by \cite{li2017deep}, which proposed to switch from \ac{CE} loss to Triplet loss towards the end of speaker embedding training for superior generalization.
On the pre-trained multi-channel model, we minimize $\mathcal{L}_{\text{triplet}} = f_\beta (d(a,p) - d(a,n) + m)$, where $f_\beta(x) = \beta^{-1} \log(1+\exp(\beta x))$, $\mathcal{L}_{\text{triplet}}$ refers to the contrastive (Triplet) loss, $f_\beta(\cdot)$ is the softplus function (smooth version of hinge function~\cite{hermans2017defense}), $\beta$ is a non-negative constant, and $d$ is the Euclidean distance function.
In the Triplet loss terminology~\cite{li2017deep}, $a$ is an \emph{anchor} example, $p$ is a \emph{positive} example, $n$ is a \emph{negative} example, and $m$ is the \emph{margin}.
$a$ and $p$ belong to the same class while $n$ belongs to a different class.
For triplet formation, we follow the \emph{hardest} mining strategy~\cite{hermans2017defense}.
It refers to choosing \emph{hardest positive} and \emph{hardest negative} example for each anchor with $d$ as the criterion.
For this, the training batch must contain multiple examples per speaker.
We follow \emph{PK sampling} i.e. choose $K$ examples per $P$ unique speakers.
We are particularly interested in \emph{large margins} in this formulation.

\section{Multi-Channel Features}
\label{sec:feats}
\underline{\textbf{Inter-Channel Phase Difference}}: \ac{IPD} is a common \emph{spatial} feature which measures the phase difference between complex \ac{STFT} of signals at two different microphones.
\begin{equation}
    {\tt{IPD}}_{(i,j), t f}=\angle\left(\frac{{Y}_{i, t f}}{{Y}_{j, t f}}\right), i,j=1\dots M, j \neq i.
\end{equation}
Here, $M$ is the number of microphones in the array, $i,j$ are microphone indices, $(t,f)$ is the current \ac{T-F} bin, and ${Y}$ is the \ac{STFT}.
Note that this gives us $M(M-1)$ pairs but we later choose only a pre-defined small subset.
Cosine and sine of \ac{IPD} are extensively used in prior works~\cite{gu2020enhancing,wang2018combining,wang2018multi} and they are referred to as cosIPD and sinIPD. 
%Note different speaker characteristics could cause different IPD even they are speaking from the same location.
%, and they are more effective than other traditional features such as \ac{GCC} features~\cite{wang2018multi}.

\underline{\textbf{Directional Power Ratio (DPR)}}: We adopt \emph{directional} features from \cite{gu2019neural} based on the output power of multi-look fixed beamformers and direction beam of target speaker $\theta_p$.
For a set of direction grid of beams $\{\theta_1,\dots,\theta_P\}$  and the corresponding filters $\mathbf{w}_{f}(\theta_p)$,
\begin{equation}
    {\tt{DPR}}_{tf}(\theta_{p})=\frac{\left\|\mathbf{w}_{f}^{\sf H}(\theta_p) \mathbf{Y}_{t f}\right\|_{2}^{2}}{\sum_{p=1}^{P}\left\|\mathbf{w}_{f}^{\sf H}(\theta_p) \mathbf{Y}_{t f}\right\|_{2}^{2}},
    \label{eq:dpr}
\end{equation}
where $\textbf{Y}$ is the mixed multi-channel complex spectrum.
The DPR features quantify how well a \ac{T-F} bin is represented by a signal from a direction beam $\theta_p$.
%DPR features are also referred to as \emph{fixed beamformer} features here.

% \underline{\textbf{Fixed Beamformer Features}}: We use a simple \emph{fixed beamforming} technique called \ac{DS}.
% To compensate for the \ac{TDOAs} on the array, different channel signals are delayed to compute the beamformed signal:
% \begin{equation}
%     y_{\tt{DAS}}(t)=\sum_{i=1}^{M} y_{i}\left(t-\delta_{i}\right)
% \end{equation}
% Here, $t$ is the time index, $y_i$ is the signal at microphone $i$, and $\delta_{i}$ is the corresponding time-difference calculated using array geometry and DOA of source.
% Finally, \ac{LPS} of $y_{\tt{DAS}}$ is considered as the \emph{fixed beamformer} features (DPR).
% %We don't experiment with adaptive beamforming~\cite{chen2018multi} that has been shown to further improve target speaker discriminability.

\underline{\textbf{Angle Features}}: We adopt the Angle Features (AF) from \cite{gu2020multi}.
% Austin: Note our AF is slightly different from the original
% Proposed in \cite{chen2018multi}
% \begin{equation}
%     A_{n, t f}=\sum_{i=1}^{M} \frac{e_{n}^{i, f} \frac{y_{i, t f}}{y_{1, t f}}}{\left|e_{n}^{i, f} \frac{y_{i, t f}}{y_{1, t f}}\right|}
% \end{equation}
The AF is formed according to the angle $\theta$ of a target speaker and measures the cosine distance between the steering vector and IPD:
\begin{equation}
\begin{aligned}
\centering
A_{tf}(\theta_t) &={\underset{(i,j)}{\sum}}
{\left < \mathbf{e}^{ {\tt TPD}_{(i,j),tf}},
\mathbf{e}^{{\tt IPD}_{(i,j),tf}} \right>
}
\\
{\tt TPD}_{(i,j),tf}&= 2\pi \varDelta_{ij}  \cos{(\theta_t)} f / (c \cdot f_s)
\end{aligned}
\label{eq:DF}
\end{equation}
Vector $\mathbf{e}^{(\cdot)}=[\cos(\cdot),\sin(\cdot)]^{\sf T}$, ${\tt TPD}_{(i,j),tf}$ (Target-dependent Phase Difference) is the phase delay between microphones $i$ and $j$ for a plane wave with frequency $f$. The plane wave is travelling from an angle $\theta_t$ (target speaker's angle at time $t$), $\varDelta_{ij}$ is the distance between the $i,j$ microphone pair, $c$ is the sound velocity and $f_s$ is  sampling rate.
%AFs are independent of microphone array topology.
% The pre-masking step in \cite{chen2018multi} is also applied to the AF to increase the discriminativity between speakers. Note that Eq. \ref{eq:DF} is reformulated so that it can be applied to general microphone array topology rather than the special seven-element microphone array used in \cite{chen2018multi}.
%\underline{How to obtain the speaker's angle $\theta$?} 
In the training stage, the angle of target speaker $\theta$ is known as multi-channel audios were generated by simulation.
In practice, the angle of the target speaker $\theta_t$ can be estimated by the face tracking system \cite{dlib09} or audio-based source localization \cite{aswin2020}. 
%One advantage of using face tracking to estimate $\theta$ is that it is invariant to acoustic environment and can achieve more robust results \cite{gu2020multi,gu2019neural}.
Details of angle estimation are beyond the scope of this work and were discussed in \cite{zhang2020}.

\underline{\textbf{Multi-Channel SincNet Features}}: A \emph{sinc} convolution layer \cite{ravanelli2018speaker} learns data-dependent 1-D filters.
%\label{sec:sinc}
We extend such layer to multi-channel input as done in multi-channel speech enhancement work of \cite{liu2020multichannel}.
We learn $C_{\text{sinc}}=257$ filters so that the resultant feature dimension is compatible with other features.
$i$-th \emph{sinc} filter $\mathbf{s}_{i,t}$ is windowed and subsequently convolved with time-domain signal of every channel and given by:
\begin{equation}
    \begin{aligned}
    \mathbf{s}_{i,t} &=2 f_{i,\text{low}} \operatorname{sinc}\left(2 \pi f_{i,\text{low}} t\right)-2 f_{i,\text{high}} \operatorname{sinc}\left(2 \pi f_{i,\text{high}} t\right) 
    \end{aligned}
\end{equation}
$f_{i,\text{low}}$ and $f_{i,\text{high}}$ are learnable low and high cutoff frequency parameters~\cite{liu2020multichannel} and $t$ is the temporal index.
We propose to use the resultant features, \emph{MultChanSinc}, as an alternative to LPS+IPD since it combines \emph{spectral} and \emph{spatial} information.
For fair comparison, we provide \emph{MultChanSinc} same channel information as the \ac{IPD}.

\section{Experiments}
\label{sec:exp}
Our study is device-specific and is performed with a 15-channel non-uniform linear microphone array with spacing 7-6-5-4-3-2-1-1-2-3-4-5-6-7 in cm and a 180-degree camera.
%The real recorded multi-channel audio visual data will be open released soon \cite{zhang2020}.
The proposed multi-channel speaker verification system was evaluated under four far-field conditions --- 1) simulated single-talker speech, 2) simulated multi-talker speech, 3) replayed recordings of multi-talker speech, and 4) real recording of multi-talker speech.

For simulated single and multi-talker setups, test data is originally single-channel 16KHz and is simulated to 15-channel. The details of the simulation can be found in the Alg.~1 of \cite{gu2020multi}. The evaluation data consists of clean internal Mandarin corpus (36 hrs, 364 speakers). For data augmentation, we use a large internal noise and \ac{RIR} corpus which contains 700 noise files and 3000 \ac{RIR}s.
The \ac{RIR}s are multi-channel signals simulated in various virtual room configurations.
They are also similarly split to create disjoint training, validation, and evaluation data.
We constrain all sound sources (target speaker, interference speaker, noise sources) to the azimuth angle range of $(0\degree,180\degree)$.
For DPR features, we choose the spatial resolution of 10\degree and hence $P=10$ in Eq.~\ref{eq:dpr}.
Ignoring spatial ambiguity issue~\cite{gu2020enhancing}, we keep the angle between the target and interference speaker unconstrained.
The position of all sources is static.
Only one interference speaker is allowed with probability $p_{\text{tar}}$.
For training and validation, \ac{SNRs} $\in \{-3,0,6,12,18\}$ dB, \ac{SIRs} $\in \{0,3,6\}$ dB, and $p_{\text{tar}}=0.15$.
For evaluation on simulated data,  we create 10,000 one-vs-one trials with equal number of target and non-target trials, with \ac{SNRs} $\in \{-2,2,4,8,10,14\}$ dB, \ac{SIRs} $\in \{2,4\}$ dB, and $p_{\text{tar}}=0.2$.

\begin{figure}[h]
%\begin{minipage}{.4\textwidth}
\centering
\resizebox{0.47\textwidth}{!}{
\includegraphics[width=0.5\textwidth]{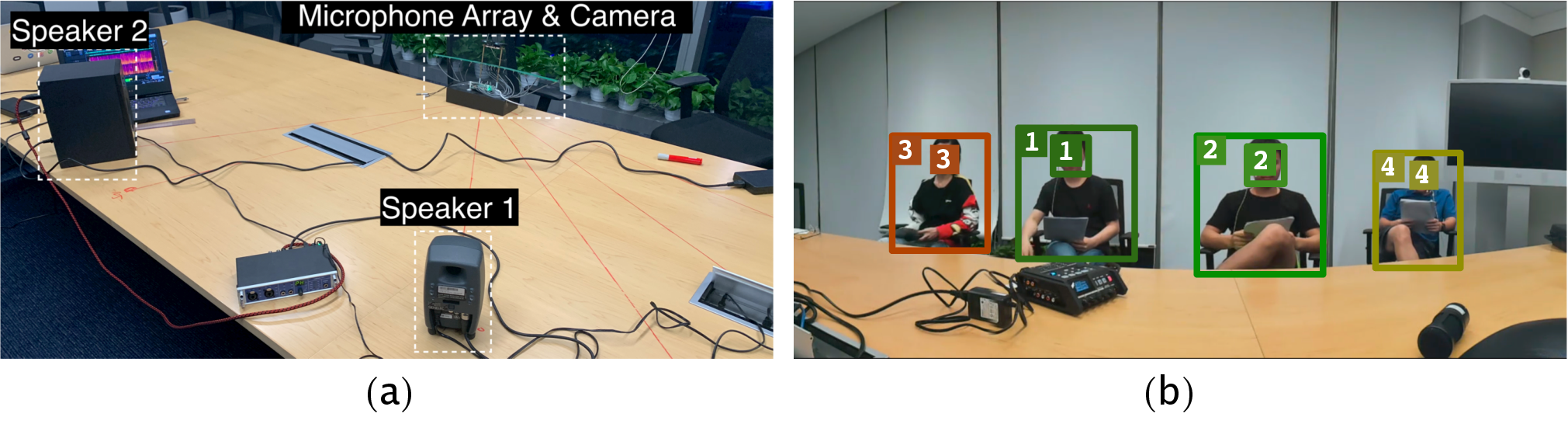}
}
%\end{minipage}%
\vspace{-0.6cm}
\caption{(a) Replayed recording of multi-talker speech. (b) Real recording of multi-talker speech with 15-channel microphone array and 180-degree camera.}
\label{fig:replay} %\vspace{-0.5cm}
\end{figure}

For the replayed multi-talker test set, 2500 (1h) utterances from 364 speakers are recorded in a 10×5×3 m$^3$ meeting room. As shown in Fig.~\ref{fig:replay}~(a), two loudspeakers are used to replay different utterances of the internal Mandarin corpus simultaneously to generate overlapped speech. We create 5,000 one-vs-one trials with an equal number of target and non-target trials for replayed data.
For the real recording of the multi-talker test set, 10h partially overlapped speech from 50 speakers in 100 debate meetings are recorded in several meeting rooms. Each speaker has recorded 3-5 single-talker utterances using the same multi-channel device, as enrollment. We create another 5,000 one-vs-one trials from segmented real recording data. This multi-channel audio-visual data of real debate meetings will be released in the near future~\cite{zhang2020}.
\vspace{-0.1cm}

\vspace{-0.25cm}
To construct training data, we first combine a portion of VoxCeleb~\cite{nagrani2020voxceleb} (450 hrs, 7285 speakers) and a small amount of internal Mandarin data (50 hrs, 2265 speakers).
This gives us approximately 500 hrs and 9550 speakers which is then split in 9:1 for training and validation data.
All \ac{STFT} based features are extracted with 512-point \ac{DFT}, 25ms analysis window, and 10ms shift.
For \ac{IPD} computation, we experiment with two sets of mic pairs:\\
\underline{v0}: {(0,7), (2,7), (3,11), (5,9), (11,5), (9,3)}\\
\underline{v1}: {(0,2), (3,5), (6,8), (9,11), (12,14), (1,4), (5,8), (9,12)}\\
%\underline{v2}: {(0,1), (3,6), (4,8), (10,14), (11,13), (0,4), (1,3), (6,10),(8,11), (13,14)}.
%\vspace{-0.25cm}
When features like DPR and AF use target speaker location only, they are referred to as ``DPR(1)'' and ``AF(1)''.
When interference speaker location is also used, we term them ``DPR(2)'' and ``AF(2)''.
When interference speaker is absent, interference noise source location is used instead.

For the speaker embedding network, we use the standard ResNet-34 architecture with \ac{TAP} of embeddings ~\cite{jung2019self}.
For the speech enhancement network, we use the multi-channel TasNet with 4 \ac{TCN} blocks. Each TCN block is stacked by 8 Dilated 1-D ConvBlock with exponentially increased dilation factors $2^0, 2^1, ....,2^7$. Each dilated 1-D ConvBlock consists of a 1$\times$1 convolutional layer, a depth-wise separable convolution layer (D--Conv) \cite{DCONV}, with PReLU \cite{PReLU} activation function, and normalization added between each two convolution layers. Skip connection is added in each dilated 1-D ConvBlock. The early fusion method \cite{gu2019Multi} was adopted to combine LPS, IPD, and AF features before feeding into a TCN block.
For \ac{VAD} network, we use a 2-layer \ac{BLSTM}.
We do Adam optimization with initial \ac{LR} of 0.0075, batch size of 240, 8 epochs, and a simple \ac{LR} scheduler (decrease by 50\% when validation loss does not improve).
%\footnote{\url{https://github.com/lessw2020/Ranger-Deep-Learning-Optimizer}}
For contrastive fine-tuning experiments, $K=4$ for \emph{PK sampling}, $P=60$ (since batch size is 240), $\beta=1$ for softplus function, and LR is 0.01 with similar scheduling as before.
Our 512-D speaker embeddings (extracted from the penultimate layer of ResNet) are compared using a simple cosine similarity backend~\cite{li2017deep} and results are reported in terms of validation accuracy, \ac{EER}, and \ac{minDCF} with the target prior probability of 0.05.
The lower the EER and minDCF, the better.

\section{Results}

\begin{table*}[htbp]
\centering
\caption{The results of multi-channel systems on four test sets, including simulated far-field single talker speech, simulated multi-talker speech, replayed recording of multi-talker speech, real recording of multi-talker speech. ``-'' marks clean testing conditions when DPR(2), AF(2) cannot be computed due to non-existence of inteference sources. \textit{Enh} and \textit{VAD} indicate if enhancement and VAD module are present respectively. \textit{val acc} (in \%) is the corresponding system validation accuracy. “DPR(2)” and “AF(2)” means the locations of both target and interference speaker are used.
}
\resizebox{1.00\textwidth}{!}{
\begin{tabular}{|c|c|c|c|c|c|c|c|c|c|c|c|}
%\hline
 \multicolumn{4}{c|}{  } & \multicolumn{2}{c|}{\cellcolor{black}\textcolor{white}{simulated (single)}} & \multicolumn{2}{c|}{\cellcolor{gray}\textcolor{white}{simulated (multi-talker)}} & \multicolumn{2}{c|}{\cellcolor{black}\textcolor{white}{replay (multi-talker)}} & \multicolumn{2}{c|}{\cellcolor{gray}\textcolor{white}{real (multi-talker)}} \\ \hline
\BlackCell{Input Features}         & \GrayCell{Enh} & \GrayCell{VAD} & \textit{val Acc} & EER (\%)      & minDCF     & EER (\%)            & minDCF           & EER (\%)          & minDCF     & EER (\%)          & minDCF      \\ \hline  \hline
{single channel LMFB}      &\xmark&\cmark            & 62.3         & 13.9          & 0.747      & 24.9                & 0.921            & 28.4              & 0.968   &  19.3  &  0.848     \\ \hline
single channel LPS                                                    &   \xmark       &\cmark&64.0& 12.7          & 0.660      & 21.5                & 0.875            & 25.0              & 0.952 & 18.6 &  0.839       \\   \hline  \toprule \hline  %{|=|=|=|=|=|=|=|=|=|=|}
MultChanSinc                                           &    \xmark      &\cmark&66.6& 12.1          & 0.654      & 23.0                  & 0.902            & 26.6              & 0.917  & 17.7 & 0.834       \\ \hline
LPS + cosIPDv0                                         &\xmark          &\cmark&66.4& \textbf{10.5}          & 0.645      & 19.9                & 0.840            &   22.2              & 0.884  & 15.2 & 0.784       \\ \hline
LPS + cosIPDv1                                         &    \xmark      &\cmark&67.2& 15.2          & 0.713      & 23.9                & 0.846            &   25.9            & 0.887 &  17.4  & 0.831         \\ \hline
%LPS + cosIPDv2                                         &    \xmark      &\cmark&67.4& 10.9          & \textbf{0.602}      & 19.2                & 0.820            &   21.8            & 0.871 &  &          \\ \hline
LPS + cosIPDv0 + DPR(1)                                &    \xmark      &\cmark&69.1& 13.7          & 0.755      & 20.0                & 0.900            & 22.1 & 0.933 &  15.0  &  0.752      \\ \hline
LPS + cosIPDv0 + AF(1)                                 &\xmark          &\cmark&71.0& 14.2          & 0.740      & 19.1                & {0.818}            & 20.7 & 0.882 &  14.5 & 0.746      \\ \hline
{LPS + cosIPDv0 + DPR(1) + AF(1)}                        & \xmark         &\cmark&70.9& 12.0          & 0.658      & {16.9}                & {0.818}            & {18.1}              & {0.866}    &  14.1 & 0.725     \\ \hline
{LPS + cosIPDv0 + DPR(1) + AF(1)}                        & \cmark         &\cmark &  70.6 &   \textbf{10.5}        &  \textbf{0.611}     &  16.4    & \textbf{0.807}  & 17.7               &  0.860   & 12.9 & {0.717}     \\ \hline
{LPS + cosIPDv0 + DPR(2) + AF(2)}                        & \cmark         &\cmark &  71.7 &     -      &  -     &   \textbf{16.3}    &  0.812    &   \textbf{17.5}            &  \textbf{0.858}   &  \textbf{12.6}&   \textbf{0.715}    \\ \hline
%\begin{tabular}[c]{@{}c@{}}LPS + cosIPDv0 + DPR(2) + AF(2) \end{tabular}                        & \cmark         &\cmark & 72.1 &   -        &  -     &   16.8   &  0.825           &   17.8            &  0.869   & &      \\ \hline
LPS + cosIPDv0 + DPR(2) + AF(2)                         & \cmark         &\xmark & 73.5 &   -        &  -     &   17.4   &  0.822           &   18.3            &  0.862   &  14.2 &  0.739    \\ \hline

%
%LPS + cosIPDv0 + DPR(1) + AF(1) + sinIPDv0             &\xmark          &\cmark&72.3& 17.5          & 0.851      & 22.0                & 0.919            &  & & &          \\ \hline
%LPS + cosIPDv0 + DPR(2) + AF(2) + sinIPDv0             &      \xmark    &\cmark&72.0& -             & -          & 19.0                & 0.891            & &&&         \\ \hline
%LPS + cosIPDv0 + DPR(2) + AF(2) + sinIPDv0 + phase0    &    \xmark      &\cmark&71.1& -             & -          & 22.2                & 0.870            & 22.6              & 0.882          \\ \hline
%LPS + cosIPDv0 + DPR(2) + AF(2) + sinIPDv0        &\cmark          &\cmark&71.6& -             & -          & 19.2                & 0.821            & &&&      \\ \hline
%LPS + cosIPDv0 + DPR(2) + AF(2) + sinIPDv0  &      \cmark    &\xmark&\textbf{73.4}& -             & -          & 18.5                & 0.88             & &&&       \\ \hline
\end{tabular}
}
\label{tab:all}\vspace{-0.3cm}
\end{table*}

\subsection{The benefit of multi-channel features over single-channel features for Speaker Verification}
In Table \ref{tab:all}, we evaluate our multi-channel system on four testing scenarios.
First, we observe that 257-D \ac{LPS} features are better than the baseline 80-D LMFBs perhaps due to higher dimensionality.
In clean conditions, \emph{MultChanSinc} features are better than single-channel features but perform worse or equivalent to LPS in noisy conditions.
This suggests that these features might not be robust to noise or are sensitive to its hyper-parameters.
LPS+cosIPD are conventional choice of features for MCSV and we find them superior to \emph{MultChanSinc} in our case.
By experimenting with two choices of microphone pairs, we observe that this choice is important and perhaps can even be learned.
%Performance-wise ranking of such pairs is v0 $>>$ v1.
We find v0 to be much superior to v1 even though the latter is a bigger set.
They also achieve the best clean condition EER.
Adding beamformer features (DPR) hurt in clean condition while roughly maintaining performance in other cases.
This is possibly because these features fail to complement other features under our simple feature fusion scheme.
The trend is similar for Angle Features (AF) although they seem superior to DPR features.
When both types of \emph{directional} features are combined, we observe complementary behavior and non-target \emph{cancellation ability}.
When speech enhancement is introduced, we observe consistent and significant gains.
Using interference source location gives us the best system which outperforms in almost all noisy conditions.
We expect this improvement to be higher when the number of sources is more than (current) two.
We also observed that sinIPD features do not deliver more information to the system (not listed here).
Finally, from the ablation experiment (last row), we can see the importance of VAD in our joint learning system.

%Incorporating sinIPD features vastly degrade performance.
%This ill-effect is perhaps also due to concatenating vastly different types of features on channel dimension and not learning to fuse them.
%However, adding interference location information gives some expected improvement.
%Including absolute phase information of the first microphone \emph{phase0} hurts, which shows that relative phase information (through IPDs) is sufficient.
%Note that, till now, enhancement is not incorporated.
%Using it further improves performance.
%However, excluding \ac{VAD} from this system degrades minDCF while improving \ac{EER}.
%This suggests that there is an interplay between enhancement and \ac{VAD} module and therefore a further exploration is required.

A critical observation is that progressively incorporating more information in our methodology leads to consistent improvement in validation accuracy, i.e. better speaker identification accuracy, but the generalization performances measured by speaker verification tasks vary.
Finally, we note that combining the LPS + cosIPDv0 + DPR + AF features has almost closed the performance gap between the real recording of multi-talker speech (12.6\%) and single-talker speech (10.5\%).

\subsection{Multi-channel system evaluation on various SNR, SIR}
In Table \ref{tab:snr}, we evaluate few key feature combinations on specific (SNR, SIR) pairs.
We note that the trend of consistent improvement with the addition of features holds true especially in challenging conditions (low SNR, low SIR).
On one cleaner condition (SIR=6, SNR=15), \emph{directional} features are worse than \emph{spatial} features.
%This explains some unexpected observations in Table \ref{tab:all} since we report averaged performance there.
This suggests the sub-optimality of our feature concatenation scheme for \emph{directional} features.
We also observe that the performance of multi-channel features is always better than single-channel and even in high SNR, high SIR conditions.
Combining \emph{spectral}, \emph{spatial}, and \emph{directional} features in low SNR=-2 dB gives almost identical performance for both SIR=0 and SIR=6.
This demonstrates the strong ability of our system to handle interference speakers.

% features $\downarrow$ \hspace{1em} \textbackslash \hspace{1em} SNR $\rightarrow$

\begin{table}\vspace{0.1cm}
\caption{The detailed results of simulated multi-talker speech on different SIRs and SNRs.}
\setlength\tabcolsep{3.5pt}
\resizebox{0.47\textwidth}{!}{
\begin{tabular}{|c|c|c|c|c|c|c|c|c|}
\hline
\cellcolor[HTML]{333333}{\color[HTML]{FFFFFF} }                                 & \cellcolor[HTML]{C0C0C0}                      & \cellcolor[HTML]{C0C0C0}                      & \multicolumn{3}{c|}{SIR=0, SNR=} & \multicolumn{3}{c|}{SIR=6, SNR=} \\ \cline{1-1} \cline{4-9} 
\multirow{-2}{*}{\cellcolor[HTML]{333333}{\color[HTML]{FFFFFF} Input Features}} & \multirow{-2}{*}{\cellcolor[HTML]{C0C0C0}Enh} & \multirow{-2}{*}{\cellcolor[HTML]{C0C0C0}VAD} & -2      & 5       & 15     & -2      & 5       & 15     \\ \hline
single channel LMFB                                                                            &  \xmark                                              &  \cmark                                              & 32.0    & 24.6    & 20.6   & 30.8    & 22.8    & 18.1   \\ \hline
single channel LPS                                                                             &   \xmark                                             &   \cmark                                             & 27.4    & 21.6    & 18.0   & 26.2    & 20.1    & 16.7   \\ \hline \toprule \hline
LPS+cosIPD                                                                      &    \xmark                                            &  \cmark                                             & 26.1    & 19.9    & 16.1   & 24.9    & 18.4    & 14.5   \\ \hline
LPS+cosIPD+DPR(1)+AF(1)                                                         &  \xmark                                             &  \cmark                                             & 23.8    & 17.1    & 15.8   & 22.1    & 17.3    & 15.4   \\ \hline
LPS+cosIPD+DPR(1)+AF(1)                                                         &  \cmark                                             &  \cmark                                             & \textbf{19.7}    & \textbf{16.5}    & \textbf{14.9}   & \textbf{19.2}    & \textbf{16.2}    & \textbf{14.1}   \\ \hline
\end{tabular}\label{tab:snr}
} \vspace{-0.5cm}
\end{table}

\subsection{Contrastive fine-tuning of the multi-channel system}

We also investigate if one of our best multi-channel systems in Table \ref{tab:all} (LPS + cosIPDv0 + DPR(1) + AF(1)) can be fine-tuned to improve verification performance by a simple choice of contrastive loss: Triplet Loss.
By experimenting with various values of margin $m$, we obtain large improvements in clean as well as noisy conditions.
This suggests that softmax training leaves some scope of improvement in embedding space.
$m=2$ gives the best results contrary to small values like 0.1 and 0.2 used in previous works~\cite{li2017deep,bredin2017tristounet}.
For simulated single and multi-talker tests, compared to the single-channel baseline of Table \ref{tab:all}, we reduce EER from 16.9\% to 15.5\%.
Hence, we demonstrate that \ac{MCSV} performance can be improved via a better design of input feature space as well as target embedding space.

\begin{table}[htbp]
\centering
\caption{Fine-tuning the multi-channel system, LPS + cosIPDv0 + DPR(1) + AF(1) with VAD, using different Triplet margin $m$.}
\resizebox{0.48\textwidth}{!}{
\begin{tabular}{c|c|c|c|c|}
                                & \multicolumn{2}{c|}{\cellcolor{black}\textcolor{white}{simulated (single-talker)}} & \multicolumn{2}{c|}{\cellcolor{gray}\textcolor{white}{simulated (multi-talker)}} \\ \cline{2-5}
                                & EER (\%)      & minDCF     & EER (\%)            & minDCF           \\ \hline
\multicolumn{1}{|c|}{no fine-tuning} & 12.0          & \textbf{0.658}      & 16.9                & 0.818            \\ \hline \toprule \hline
\multicolumn{1}{|c|}{$m=0.3$}            & 11.0          & 0.705      & 16.8                & 0.813            \\ \hline
\multicolumn{1}{|c|}{$m=1$}              & \textbf{10.2}          & 0.697      & 16.4                & 0.834            \\ \hline
\multicolumn{1}{|c|}{$\mathbf{m=2}$}              & 10.5          & 0.680      & \textbf{15.5}                & \textbf{0.780}            \\ \hline
\multicolumn{1}{|c|}{$m=3$}              & 11.3          & 0.693      & 16.4                & 0.827            \\ \hline
\end{tabular}
}\vspace{-0.5cm}
\end{table}

\section{Conclusion}
To advance robust location-aware multi-channel Speaker Verification, we tackle the problem of designing better input features as well as embedding space.
For the former task, we explored various combinations of \emph{spectral}, \emph{spatial}, and \emph{directional} features to find that single-channel baseline can be vastly improved under all testing conditions with combinations of \emph{spectral} and \emph{spatial} features.
We find the benefit of adding \emph{directional} features more prominent in multi-talker conditions while an analysis showed that further exploration is required to improve our simple feature concatenation scheme.
Overall, we observe 36\% relative reduction in Equal Error Rate (EER) on real recordings.
We also show that the discriminativity of speaker embedding space can be significantly improved via a contrastive loss-based fine-tuning of our multi-channel system.
In the future, we can investigate (1) feature fusion learning schemes; (2) robustness to sound source(s) location information (modality robustness problem~\cite{gu2020multi}); and (3) explore more multi-channel features like Inter-channel Convolution Differences (ICD) \cite{gu2020enhancing}.

\clearpage

\bibliographystyle{IEEEtran}

\bibliography{template}

% \begin{thebibliography}{9}
% \bibitem[1]{Davis80-COP}
%   S.\ B.\ Davis and P.\ Mermelstein,
%   ``Comparison of parametric representation for monosyllabic word recognition in continuously spoken sentences,''
%   \textit{IEEE Transactions on Acoustics, Speech and Signal Processing}, vol.~28, no.~4, pp.~357--366, 1980.
% \bibitem[2]{Rabiner89-ATO}
%   L.\ R.\ Rabiner,
%   ``A tutorial on hidden Markov models and selected applications in speech recognition,''
%   \textit{Proceedings of the IEEE}, vol.~77, no.~2, pp.~257-286, 1989.
% \bibitem[3]{Hastie09-TEO}
%   T.\ Hastie, R.\ Tibshirani, and J.\ Friedman,
%   \textit{The Elements of Statistical Learning -- Data Mining, Inference, and Prediction}.
%   New York: Springer, 2009.
% \bibitem[4]{YourName17-XXX}
%   F.\ Lastname1, F.\ Lastname2, and F.\ Lastname3,
%   ``Title of your INTERSPEECH 2021 publication,''
%   in \textit{Interspeech 2021 -- 20\textsuperscript{th} Annual Conference of the International Speech Communication Association, September 15-19, Graz, Austria, Proceedings, Proceedings}, 2020, pp.~100--104.
% \end{thebibliography}

\end{document}